# Analytical modeling and experimentally optimizing synergistic effect on thermal conductivity enhancement of polyurethane nanocomposites with hybrid carbon nanofillers

Amir Navidfar*, Levent Trabzon

Faculty of Mechanical Engineering, Istanbul Technical University, Istanbul, Turkey;
Nanotechnology Research and Application Center, Istanbul Technical University, Istanbul, Turkey;
MEMS Research Center, Istanbul Technical University, Istanbul, Turkey;

## Abstract

Combining various carbon nanofillers with different dimensions can lead to a synergistic effect through the formation of an efficient conductive network. Hybrid polyurethane (PU) nanocomposites containing multi-walled carbon nanotubes (MWCNTs) and graphene nanoplatelets (GNPs) were fabricated to study experimental and theoretical aspects of thermal conductivity (TC) enhancement. The optimization of hybrid nanofillers combinations was done to synergically enhance the TC using various types of GNPs, different nanofillers concentrations and varied ratios. A synergistic thermal conductivity improvement with MWCNTs and GNPs was confirmed at low nanofillers contents. The TC of hybrid nanocomposite at 0.25 wt% is approximately equivalent to the TC of individual nanofillers at 0.75 wt%. An analytical model for the effective thermal conductivity of single and hybrid nanocomposites was considered with variables of volume fraction, interfacial thermal resistance, straightness of the nanofillers and the percolation effect, in which the predictions of the modified models agree with the experimental results.

***Correspondence to: Amir Navidfar (Email: Navidfar@itu.edu.tr)**

## 1. Introduction

Lightweight thermally conductive polymers have attracted a great deal of attention in some fields, such as electronic packaging, heat exchangers and flexible electronics, in which the low thermal conductivity is one of the main technical limitations [1-3]. Polyurethane (PU) foams were considered due to their low density, high global volume production, low cost and solid nanofillers inclusion facility.

Polymeric foams thermal conductivity is mainly transferred by the solid fraction and the gas encompassed in its cells controls thermal insulating properties [4, 5]. To enhance the thermal conductivity (TC) of polymer composites, introducing high thermally conductive nanofillers is a simple and common technique. Nanocarbon fillers owing to their lightweight, high thermal conductivity and high aspect ratio are receiving significant attention [6-12]. Among them, graphene nanoplatelets (GNPs) and carbon nanotubes (CNTs) due to their two-dimensional (2D) and one-dimensional (1D) structure and ultrahigh surface areas offer exceptional thermal conductivity to the polymer nanocomposites [13]. Thermal conductivities of CNTs and GNPs have been informed to reach in the range of 1950–5000 and 3000–6500 W/m K, respectively [14-18]. Recent experimental results indicate that combining two different nanocarbon fillers with varied sizes as hybrid fillers could lead to a synergistic effect because of a three-dimensional (3D) thermally conductive network formation, which exceeds the thermal properties of the individual GNPs or CNTs filled nanocomposites [17, 19-25]. Yu et al. [26] achieved a significant thermal conductivity enhancement for GNPs and single-walled CNTs hybrid fillers with a weight ratio of 3:1. He et al [20] reported that the thermal conductivity was obviously enhanced with 3D hybrid GNPs/CNTs inclusion in the polymer matrix at very low nanofiller loading. Enhancing thermal conductivity at low nanofiller loadings is essential for improving processability and reducing the viscosity and the cost of fabricated nanocomposites. On the other hand, several analytical predictions on the thermal conductivity of carbon nanofillers reinforced nanocomposites have been proposed [6, 14-16, 27]. Chu et al [16] theoretically analyzed the thermal conductivity of GNPs nanocomposites with considering the flatness influence of GNPs. Their experiments indicated that the waviness of graphene considerably affects the thermal conductivity of nanocomposites. In other work, Chu et al. [6] developed a theoretical model for anticipating the effective thermal conductivity (ETC) of hybrid CNT/GNP nanocomposites with incorporating the percolation effect. Although there are a few studies on the thermal conductivity improvement of single or hybrid carbon nanofillers based nanocomposites, simultaneously theoretical modeling and experimental studies on the thermal conductivity of individual and hybrid nanocomposites are still lacking.

In this work, we compared three types of graphene nanoplatelets with different physical properties e.g. flakes size, aspect ratio and specific surface areas to investigate graphene type dependence of hybrid GNPs/MWCNTs nanofillers on synergistic thermal conductivity of PU nanocomposites. Synergistic thermal conductivity improvement is observed by the formation of a 3D conductive structure by means of GNP-S750 with a smaller flake dimension and a higher specific surface area at a low content of 0.25 wt%. A theoretical study on the ETC of single and hybrid nanocomposites by incorporating the straightness factor and percolation effect as fitting parameters is presented. Our analytical results are found to be in reasonable agreement with experimental data, which well describe the thermal conductivity improvement of single and hybrid nanocomposites.

## 2. Experimental

### 2.1 Materials and Sample preparation

As suggested by the manufacturer (Table 1), polyurethane fabrication was carried out by blending the polyol and isocyanate with a weight ratio of 1:1.25. MWCNTs and GNPs were bought from Nanografi Co.Ltd, which MWCNTs (purity ≥ 92%) were grown by chemical vapor deposition. The properties and geometries of used MWCNTs and GNPs were listed in Table 2, according to the manufacturer datasheet and the density of MWCNTs and GNPs are 2.1 and 2.25 g/cm$^3$, respectively.

**Table 1**
Properties of polyol and isocyanate components.

| Physical properties | Unit | Polyol | Isocyanate | Standards |
|---|---|---|---|---|
| **Density (25°C)** | g/cm$^3$ | 1.11 | 1.23 | DIN 51 757 |
| **Viscosity (25°C)** | MPa.s | 600 ± 200 | 210 | ASTM D4878-98 |
| **OH content** | Mg KOH/g | 300 | - | ASTM D 4274-99 |
| **NCO content** | $H_2O$ | - | %30.8 -%32 | ASTM 5155-01 |
| **Storage life** | Month | 3 | 6 | - |

**Table 2**
Properties of carbon nanofillers.

| Nanofiller | Diameter (nm) | Thickness/Length (nm) | Specific Surface Area (m$^2$/g) |
|---|---|---|---|
| **MWCNT** | $D_{CNT}$ = 8-10 | L =1000-3000 | 290 |
| **GNP-L150** | $D_{GNP}$ = 24000 (Large) | t = 6 | 150 |
| **GNP-M150** | $D_{GNP}$ = 5000 (Medium) | t = 6 | 150 |
| **GNP-S750** | $D_{GNP}$ = 1500 (Small) | t = 3 | 750 |

Before the PU synthesis, used MWCNTs were functionalized with hydrogen peroxide ($H_2O_2$) as reported elsewhere [21, 22, 28]. Different contents and ratios of MWCNTs and GNPs were first mixed with polyol at 200-2000 rpm for 5 min using overhead stirrer equipment. Subsequently, the combination was ultrasonically dispersed for 5 min using ultrasonication and stirred again at 2000 rpm about 5 min. The isocyanate was finally added to the carbon nanofiller/polyol composition, which was stirred for 20 s and then poured into a two-part wooden mold with a diameter of 180 mm to form free-rise PU foam (Fig. 1) [21, 22, 28]. Table 3 summarizes the fabricated nanocomposites with MWCNTs, GNPs, and their compositions. For hybrid nanocomposites, the ratio of carbon nanotubes and graphene nanoplatelets was varied for each GNPs type to investigate their effectiveness in the thermal conductivity of the nanocomposites.

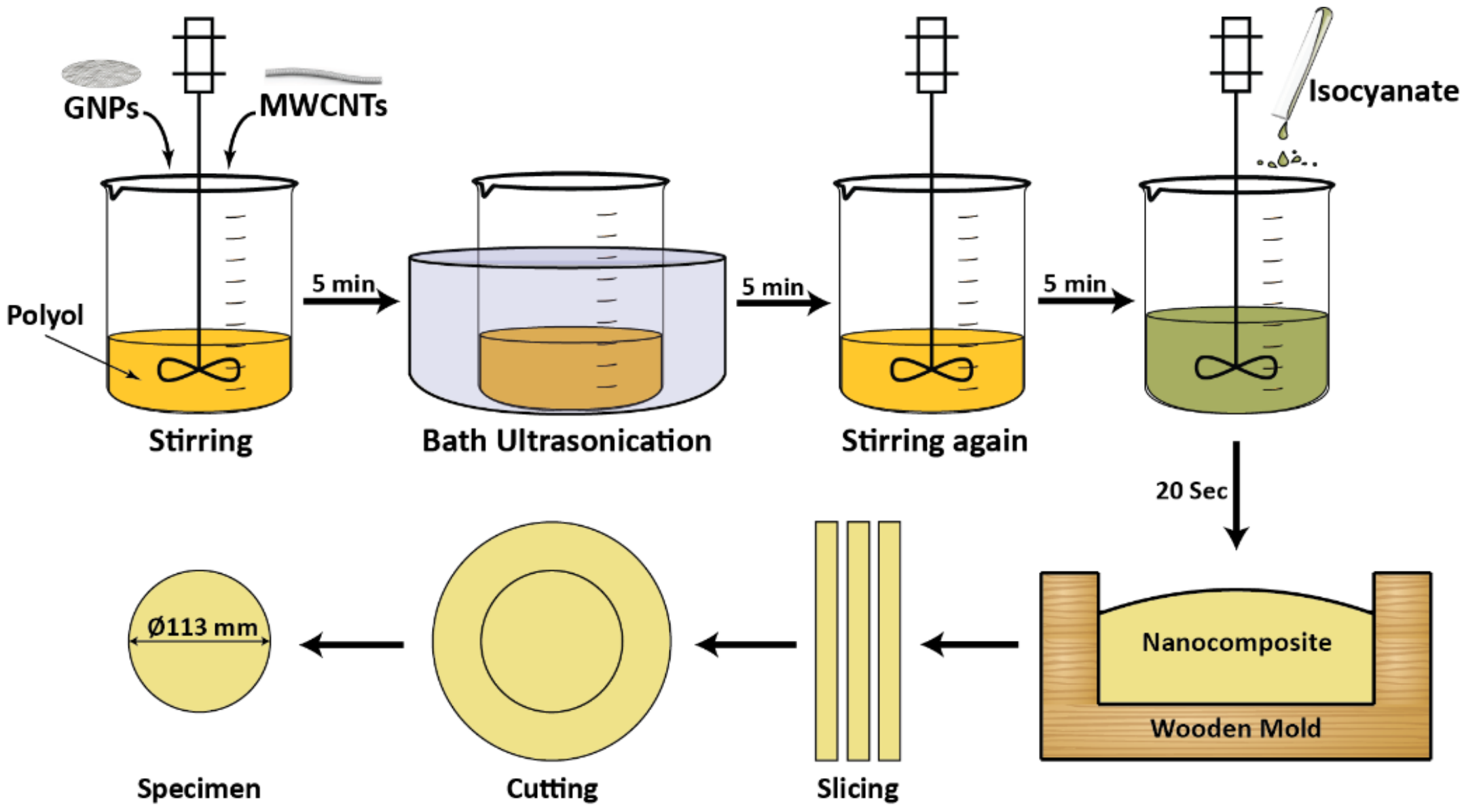

Fig. 1. The scheme of nanocomposites fabrication steps.

**Table 3**
Levels of fabricated nanocomposites.

| **Nanofillers** | **Levels (wt %)** | | |
|---|---|---|---|
| **MWCNT** | 0.25 | 0.50 | 0.75 |
| **GNP-L150** | 0.25 | 0.50 | 0.75 |
| **GNP-M150** | 0.25 | 0.50 | 0.75 |
| **GNP-S750** | 0.25 | 0.50 | 0.75 |
| **MWCNT/GNP-L150** (1:3), (1:1), (3:1) | 0.25 | 0.50 | 0.75 |
| **MWCNT/GNP-M150** (1:3), (1:1), (3:1) | 0.25 | - | - |
| **MWCNT/GNP-S750** (1:3), (1:1), (3:1) | 0.25 | - | - |

## 2.2 Characterizations

For cutting slices of cured nanocomposites with a thickness of 10 mm, a lathe machine was used. A cylindrical metal piece with sharp edges and a diameter of 113 mm were pressed over sliced PU foams to shaping thermal measurement specimens. The thermal conductivity analysis was carried out using the guarded comparative-longitudinal heat flow technique according to ASTM-E1225 test method. X-ray diffraction (XRD) analysis was determined on a RIGAKU Diffractometer using Cu (Ka) radiation. Structure analysis of carbon nanofillers was studied by means of a Renishaw inVia Raman spectrometer with a 532 nm laser. Morphologies of the nanocomposites composed of PU, GNPs and MWCNTs were observed by FEI Nova NanoSEM scanning electron microscopy (SEM). HighTech HT7700 transmission electron microscopy (TEM) was used to investigate dispersion states of carbon nanofillers and their hybrids.

# 3. Results and Discussion

## 3.1 Morphology and Characterization

Fig. 2(a) and (b) display XRD patterns of carbon nanofillers, PU and PU/GNP-S750+CNT nanocomposites in the range of $2\theta$ =10-50˚. An intense peak at $2\theta$ = 26.4° is obvious for all GNPs, whereas GNP-L150 and GNP-M150 have a much sharper peak, representing that GNP-L150 and GNP-M150 are more crystalline than MWCNTs and GNP-S750. The XRD spectra of MWCNTs exhibits a diffraction peak at 25.8˚ (002) and 43.3˚ (100), corresponding to an inner-walls distance in the radial direction of CNTs [29]. Moreover, the pristine MWCNTs and functionalized MWCNTs (f-MWCNT) demonstrate the same XRD spectra, proving that the functionalization by $H_2O_2$ did not alter the crystalline structure of CNTs (Fig. 2(b)), in contrast to the results of acid treatment such as $HNO_3/H_2SO_4$ [30]. A broad amorphous peak of PU was seen around 19˚, which confirms an amorphous microstructure of polyurethane (Fig. 2(b)). In PU/GNP-S750+CNT hybrid nanocomposite, the broad diffraction peak of the polymer matrix can be observed, while the peaks assigned to MWCNTs and GNP-S750 were disappeared, demonstrating that the carbon nanofillers were well dispersed in the PU matrix at 0.25 wt% content [31]. Raman spectroscopy is carried out to characterize crystal structure, disorders and defects in carbon-based materials. Fig. 2 (c) illustrates the characteristic peaks of carbon nanofillers, including D-band, G-band and 2D-band, which are relevant to defects and carbon structures. The D-band to G-band intensity ratio of GNP-S750 ($I_D/I_G$ = 0.49) shows a higher value compared to GNP-L150 and GNP-M150 ($I_D/I_G$ = 0.08), indicating more defects and porous graphene. This can be ascribed to the lower flake size of GNP-S750 and more functional groups on the GNP-S750 surface and edges, which increased the degree of disorders [32]. XRD analysis that revealed the higher crystalline structure of GNP-M150 and GNP-L150 supported this statement. The $I_D/I_G$ value of MWCNTs ($I_D/I_G$ = 1.19) is higher than those GNPs, which is attributed to $sp^3$ defects and twist structure of carbon nanotubes [33].

TEM micrographs of GNPs and MWCNTs and their hybrid are shown in Fig. 2 (d-f), where the wrinkled or crumpled structure of GNP nanosheets can be observed. In MWCNT/GNP-S750 hybrid (Fig. 2 (f)), MWCNTs interconnected adjacent GNPs to form a 3D hybrid network structure, which inhibited the aggregation of GNPs and CNTs. Thus, the hybrid MWCNT/GNP-S750 combination is anticipated to provide an outstanding improvement in the mechanical and thermal properties of nanocomposites [22, 34, 35].

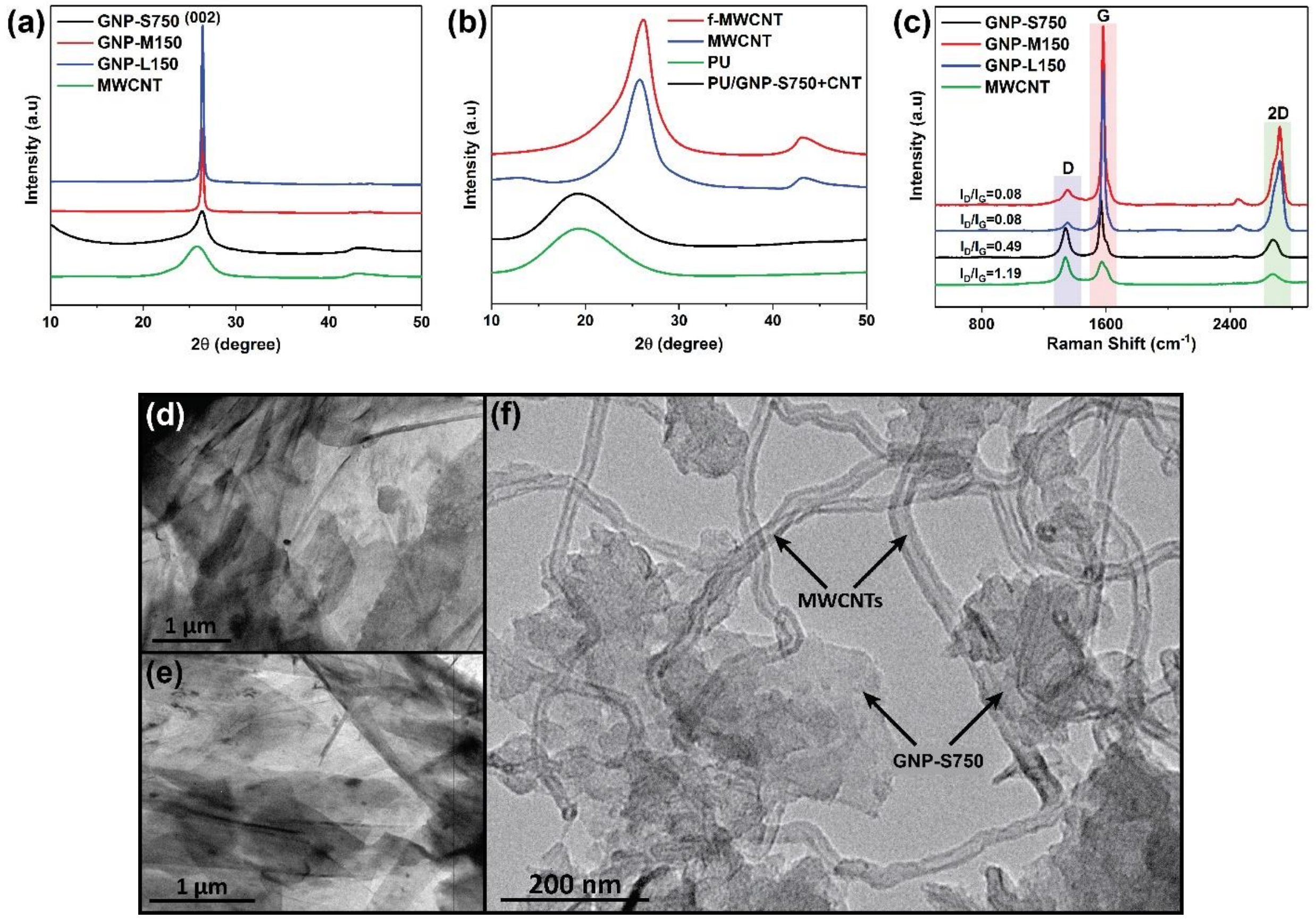

Fig. 2. (a) and (b) XRD patterns and (c) Raman spectra of carbon nanofillers and TEM images of (d) GNP-L150, (e) GNP-M150 and (f) GNP-S750/MWCNTs hybrids (an efficient 3D conductive network).

Samples cross-section were cryogenically fractured and coated with gold to observe the surface of nanocomposites. Fig. 3 demonstrates SEM images of PU nanocomposites with different carbon nanofillers, as well as hybrid GNP-S750/MWCNTs inclusion at low content of 0.25 wt%. Several aggregations are evident in PU/MWCNTs [7, 21, 36, 37], while a better nanofiller distribution in hybrid GNP-S750 and MWCNTs nanocomposites is obvious, according to Fig. 3 (e). 1D MWCNTs interconnected to 2D GNPs surface, which formed a 3D structure and hindered agglomerations [2, 38]. This 3D architecture improves the contact surface areas among MWCNTs/GNP-S750 structures and the polymer matrix that is promising for heat transferring. The folded shape of GNP-L150 with a larger flake size is clear in Fig. 3 (b), which weaken the thermal conductivity of GNP-L150 in the polymer matrix.

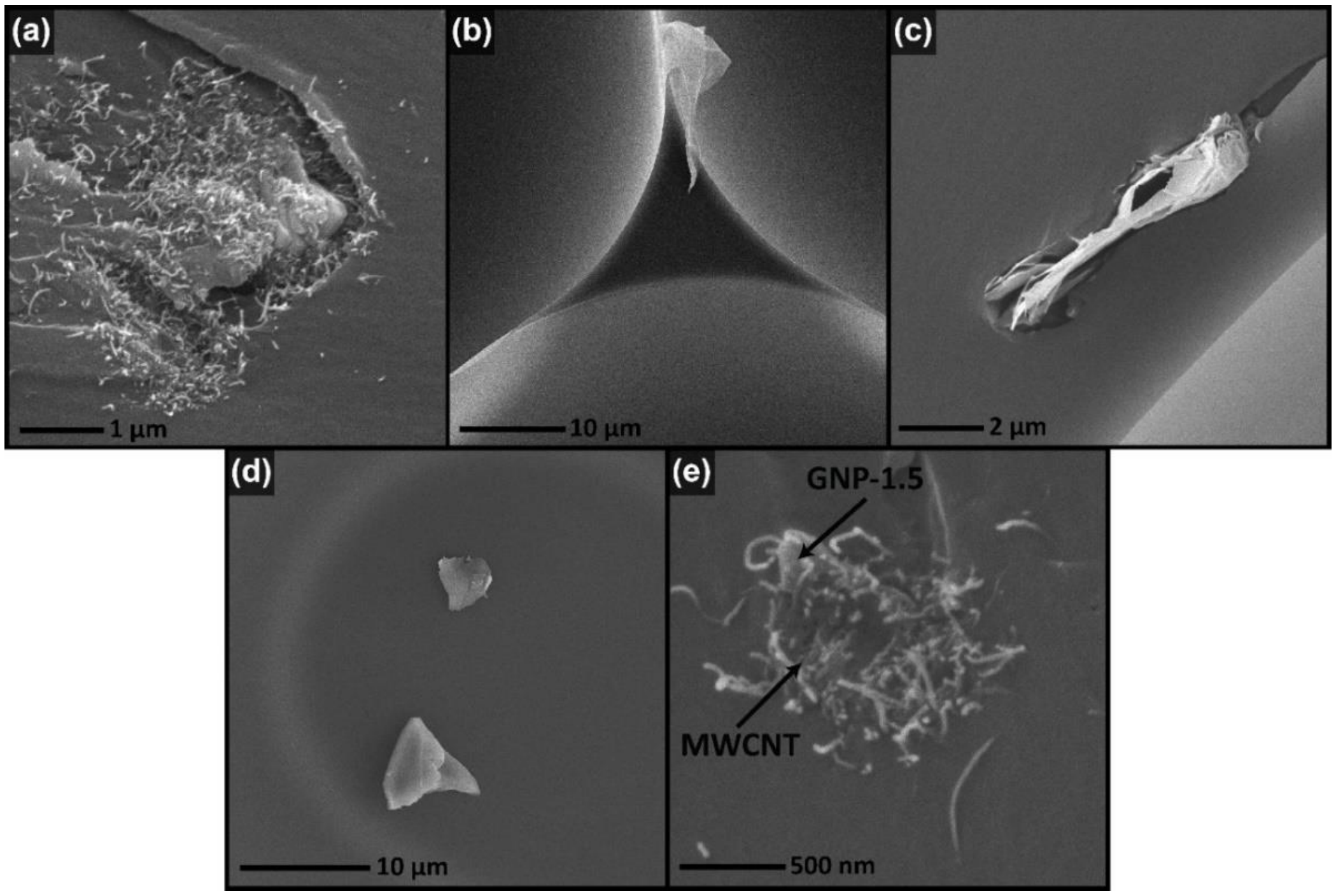

Fig. 3. SEM images of (a) PU/MWCNT, (b) PU/GNP-L150, (c) PU/GNP-M150, (d) PU/GNP-S750, (e) PU/GNP-S750 + MWCNT (1:1) nanocomposites. (0.25 wt%)

## 3.2 Thermal Conductivity

Comparative thermal conductivity results of MWCNTs and three types of GNPs in various nanofiller contents are illustrated in Fig. 4(a). Pure PU has an extremely low thermal conductivity of 0.02617 W/mK. All PU nanocomposites show the gradually enhanced thermal conductivity with increasing nanofiller loading. The highest thermal conductivity value was measured at 0.279 W/mK for 0.75 wt% of PU/GNP-S750 with a 6.6% improvement as compared to PU. The considerable improvement in thermal conductivity of PU/GNP-S750 at higher content (0.75 wt%) reflects the well-dispersed GNP-S750 in the PU matrix, due to their higher specific surface area and smaller flake size [22]. A uniform nanofiller dispersion is vital for enhancing the thermal conductivity of nanocomposites because a homogeneous nanofiller network hinders high GNP-PU interface resistance and promote phonon transport [39]. Higher surface area causes improved heat transfer and minus thermal resistance [40]. Although GNPs with a higher aspect ratio showed better thermal conductivity in the polymer matrix, the higher waviness of GNPs with a larger aspect ratio limited their conductive properties [16, 40, 41]. Thus, wrinkled GNPs would increase the interfacial bonding and decreased effectiveness of GNPs as thermal conductors in the polymer matrix [16]. Therefore a balance between nanofiller dimensions and the dispersion rate is essential in thermal conductivity. In addition, Deng et al. [14] declared that the enhancement trend of TC with increasing aspect ratio tends to be saturated for higher than 500 using an analytical model. As a result, at low nanofiller loadings (<1 wt%), there is no major difference between carbon nanofillers in the thermal conductivity of PU nanocomposite [33], according to Fig. 4(a).

Three types of GNPs were used to study the influence of graphene dimensions, specific surface area and their defects on the thermal conductivity of hybrid PU nanocomposites. Fig. 4(b) illustrates a comparative outcome of GNPs on the thermal conductivity of hybrid nanocomposites at a constant nanofiller loading of 0.25 wt%. The thermal conductivity of the nanocomposite with hybrid MWCNT/GNP-S750 nanofiller is dramatically enhanced in a synergistic manner compared to other nanocomposites. The thermal conductivity of nanocomposites with hybrid MWCNT/GNP-S750 is enhanced synergically up to 5% relative to that of PU, whereas PU/GNP-S750 and PU/MWCNT nanocomposites show 2.44% and 2.66% thermal conductivity improvement, respectively. It is clear that GNPs with a higher SSA and more defects have a higher ability to form a more efficient 3D thermally conductive network with MWCNTs and can form a path for heat flowing [22]. As an interesting outcome, the thermal conductivity of MWCNTs/GNP-S750 (1:1) hybrid nanocomposite with a synergistic effect at 0.25 wt% is approximately equal to the thermal conductivity of individual GNPs and MWCNTs at 0.75 wt% (Fig 4 (a) and (b)). Therefore, we can conclude that hybrid nanocomposites fabrication improves processability and reduces the viscosity and the cost of nanocomposites by means of declining required nanofillers concentration.

On the other hand, GNP-L150 was used to prepare hybrid nanocomposites with MWCNT in higher contents and varied ratios to study the synergistic effects of both nanofillers. The thermal conductivity of PU with MWCNT and GNP-L150 with constant loadings (0.25, 0.5 and 0.75 wt%) are shown in Fig. 4(c). A synergistic effect was not seen at 0.25 wt% and 0.5wt% nanofiller contents, whereas the synergy was obtained at 0.75 wt%. The CNTs in hybrid nanofillers act as interconnectors among GNPs and provide further channels for the heat flowing due to their high aspect ratio. The gap among GNPs is higher in lower nanofiller loadings and consequently, CNTs cannot interconnect adjacent GNPs. Our previous work showed that the synergistic effect in hybrid nanocomposites with larger GNPs can be obtained by means of CNTs with a higher length or enhancing the content and ratio of GNPs [22].

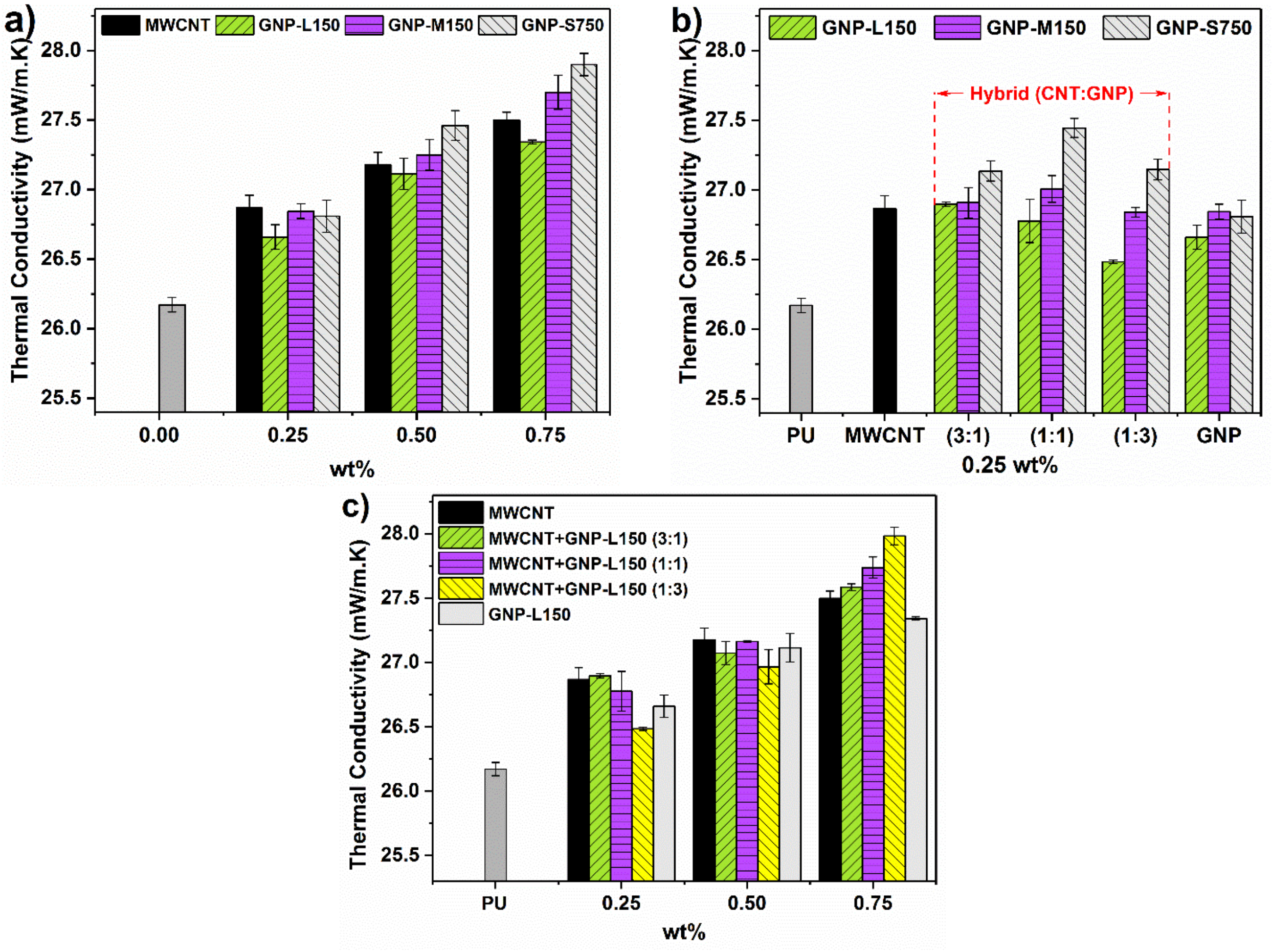

Fig. 4. The thermal conductivity of PU with (a) single nanofillers, (b) and (c) hybrid nanofillers.

## 3.3 Analytical Modeling

### 3.3.1 Single Nanofillers

To profoundly understand the thermal conductivity behavior of nanocomposites, a theoretical analysis was considered. Some factors limit the thermal conductivity of nanocomposites. The nanofiller-matrix interface resistance ($R_K$) has a negative influence on thermal conductivity of the nanocomposites, which ETC of CNTs and GNPs is given by [42]:

$$K_{CNT}^{eff} = \frac{K_{CNT}}{\frac{2R_K K_{CNT}}{L_{CNT}} + 1}, \qquad K_{GNP}^{eff} = \frac{K_{GNP}}{\frac{2R_K K_{GNP}}{L_{GNP}} + 1} \tag{1}$$

Where $L_{GNP}$ and $L_{CNT}$ are the width and length of the GNPs and MWCNTs, respectively and *eff* express ETC. The interfacial thermal resistance, $R_K$, is taken as 8 x $10^{-8}$ $m^2$ K/W for both MWCNTs and GNPs in this work [14, 42]. Microscopic analysis reveals that GNPs and CNTs in nanocomposites are frequently far from straightness owing to their large aspect ratios. The CNTs are usually crooked and GNPs are commonly wrinkled and folded in the matrix, which lead to a reduction in the ETC of nanocomposites [14-16]. The straightness factor is presented as:

$$\eta = \frac{L^e}{L} = \frac{K^e}{K} \tag{2}$$

Where L and $L^e$ are the original length and equivalent length of GNPs or CNTs, respectively. K and $K^e$ are the intrinsic and equivalent TC of GNPs or CNTs, respectively. Chu et al. [16] developed an analytical model for ETC of GNP nanocomposites using an effective medium theory of Nan et al. [42]. On the other side, a simple model for TC enhancement of CNT based nanocomposites was presented by Deng et al. [14]. Moreover, the waviness of incorporated CNTs and GNPs are found to be a significant factor in the thermal conductivity [32] and thus straightness ratio (η) are added to formulas in this work, as follow:

$$\frac{K_e}{K_m} = 1 + \frac{1}{3}\frac{\eta f_{CNT}}{H(\eta p) + K_m/\eta K_{CNT}^{eff}} \tag{3}$$

$$\frac{K_e}{K_m} = 1 + \frac{2}{3}\frac{\eta f_{GNP}}{H(\eta p) + (\eta^2 K_{GNP}^{eff}/K_m - 1)^{-1}} \tag{4}$$

H(p) is a function of nanofillers aspect ratio (p) that is negligible due to the high aspect ratio of the nanofillers and can be ignored [14]. To predict nonlinear thermal conductivity enhancement of the nanocomposites, a simple percolation model can be applied, which is described as $K_e \sim (f - f_p)^{\alpha}$. $f_p$ indicates the critical volume fraction for the nonlinear conductivity variation, and α is a critical conductivity exponent, which depends on the contact among nanofillers and/or their intrinsic conductivity. It is obvious that the measurements and prediction deviation without considering the percolation effect occurred at very low nanofiller concentrations, implying that $f_p$ may be very low and is estimated $f_p$=0.0001 [6, 20]. By incorporating the nonlinear influence of nanofillers, the final $K_e$ of the nanocomposites can be obtained:

$$\frac{K_e}{K_m} = 1 + \frac{1}{3}\frac{\eta (f_{CNT} - 0.0001)^{\alpha}}{K_m/\eta K_{CNT}^{eff}} \tag{5}$$

$$\frac{K_e}{K_m} = 1 + \frac{2}{3}\frac{\eta (f_{GNP} - 0.0001)^{\alpha}}{(\eta^2 K_{GNP}^{eff}/K_m - 1)^{-1}} \tag{6}$$

Eqs. (5) and (6) present the thermal conductivity of the nanocomposites as a function of aspect ratio, nanofiller content, anisotropy, interfacial thermal resistance, straightness factor and interaction among nanofillers. Thermal conductivity of pure PU measured $K_m$= 0.02617 W/mK and thermal conductivity of GNPs and MWCNTs regarded as 2000 W/mK [14, 16, 42]. The comparison between experimental measurements and theoretical calculations is shown in Fig. 5. Our models provide the best fit for experimental data by taking α and η as fitting parameters for various nanocomposites with nonlinear TC behavior (Table 4). The predicted values of $K_e$ using

Eqs. (3) and (4) versus weight fraction of nanofillers are plotted in Fig. 5, as well as the nonlinear prediction by Eqs. (5) and (6). As shown in Table 4, the amount of R-square is higher for Eqs. (5) and (6) using the percolation effect, compared to Eqs. (3) and (4). The results clearly show that larger GNPs tend to be wavier in the polymer matrix, so that calculated straightness factor (η) reduced by enhancing the flake size of GNPs, as presented in Fig. 5. The waviness factor (η) is equal to 0.15, 0.27 and 0.42 for PU nanocomposites with GNP-L150, GNP-M150 and GNP-S750, respectively. As a result, the straightness ratio has a dominant effect on the TC of GNPs based nanocomposites [18].

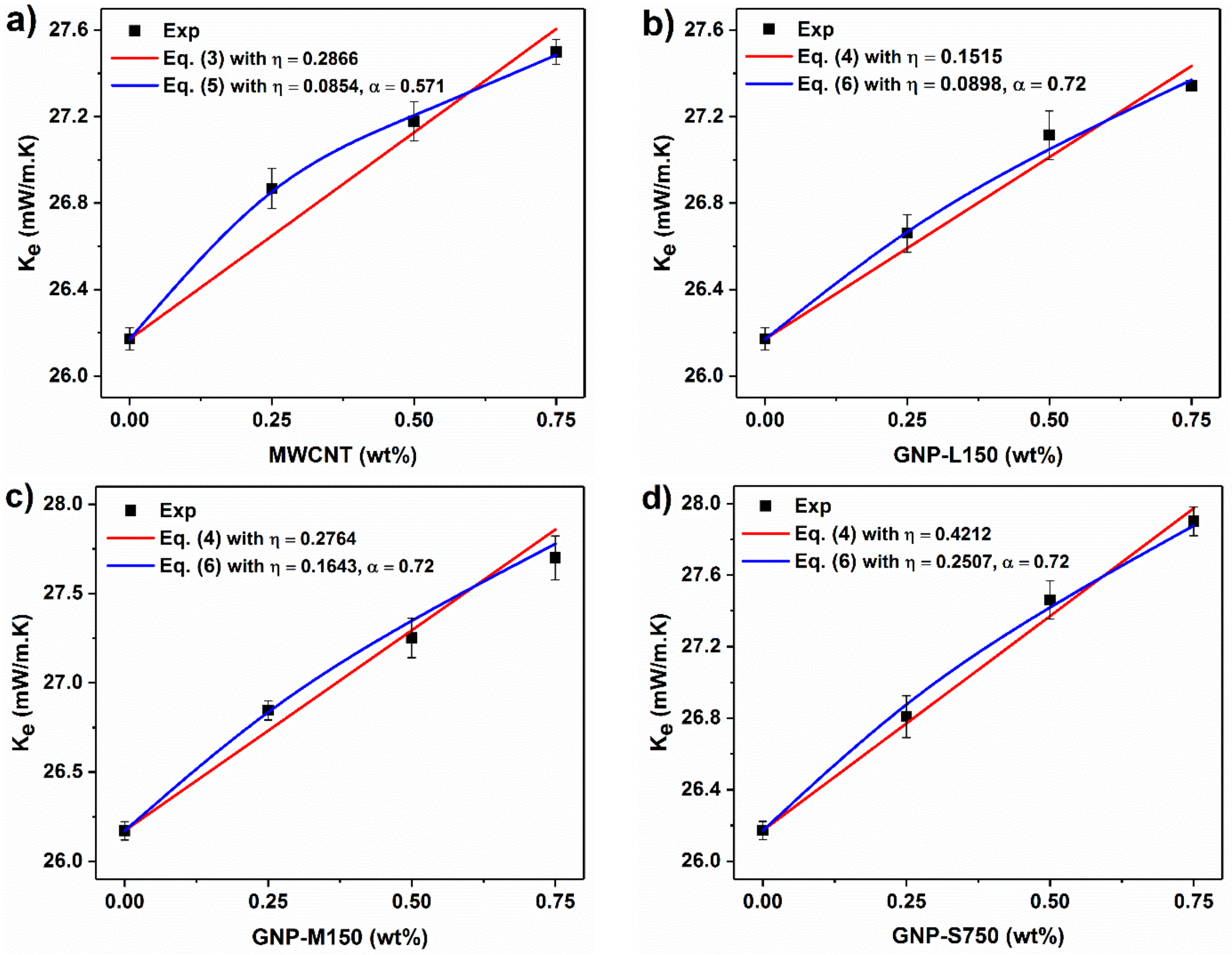

Fig. 5. Measured and theoretical modeling of (a) PU/MWCNT, (b) PU/GNP-L150, (c) PU/GNP-M150 and (d) PU/GNP-S750 nanocomposites.

**Table 4**
Fitting parameters of the analytical models.

| Carbon Nanofillers | η | α | $R^2$ (%) | Equations |
|---|---|---|---|---|
| **MWCNT** | 0.2866 | - | 84 | (3) |
| | 0.0854 | 0.571 | 99.7 | (5) |
| **GNP-L150** | 0.1515 | - | 93.8 | (4) |
| | 0.0898 | 0.72 | 98.3 | (6) |
| **GNP-M150** | 0.2764 | - | 87.1 | (4) |
| | 0.1643 | 0.72 | 97.5 | (6) |
| **GNP-S750** | 0.4212 | - | 98.3 | (4) |
| | 0.2507 | 0.72 | 98.6 | (6) |
| **MWCNT / GNP-S750*** | 0.34 | 0.89 | 100 | (7) |
| **MWCNT + GNP-S750**** | 0.34 | 0.75 | 99.9 | (7) |
| **MWCNT / GNP-L150*** | 0.27 | 1.22 | 100 | (7) |
| **MWCNT + GNP-L150**** | 0.27 | 1.33 | 99.2 | (7) |

*Individual nanofillers **Hybrid nanofillers

### 3.3.2 Hybrid Nanofillers

To well describe the synergistic effect on the thermal conductivity enhancement of PU with hybrid GNPs/CNTs, the experimental data are compared with the theoretical values. Chu et al. [6] proposed a theoretical model to describe the synergistic effect in thermal conductivity improvement of hybrid nanocomposites that is used to predict TC of hybrid GNPs/CNTs based PU nanocomposites in this work. As described before, the straightness factor of nanofillers, η, has a significant influence on the thermal conductivity of nanocomposites, which is added to the formula, as follow:

$$\frac{K_e}{K_m} = \frac{1+\eta^2 f_c(K_{CNT}^{eff}/K_m)/3 + 2\eta^3 (f_g - 0.0001)^a (K_{GNP}^{eff}/K_m)/3}{1-(2f_{CNT}+f_{GNP})/3} \quad (7)$$

It is noted that MWCNTs and GNPs are supposed to have the same η and $R_K$. The theoretical model by Eq. (7) as a function of GNPs/CNTs ratios is shown in Fig. 6, in which the experimental results are significantly matched with predicted data. The thermal conductivity of PU with hybrid GNP-S750/MWCNTs is synergically higher than those of PU with single GNP-S750 or MWCNTs at 0.25 wt%, whereas the synergy was obtained at 0.75 wt% for MWCNT/GNP-L150. The bridging interaction of GNPs and CNTs could lead to the development of a 3D thermally conductive structure that efficiently transfers the heat flow and inhibits the aggregations, as shown in Fig. 3(f). The factor α represents the nanofillers dispersion in the polymer matrix, whereas the higher α value reflects the poor dispersion of nanofillers in the polymer matrix [43]. It is determined that the fitting parameters of α=0.75 and η=0.34 for hybrid GNP-S750/MWCNTs nanocomposites at $f_{GNP} < 0.125$ wt% and α=0.89 and η=0.34 for PU with individual GNP-S750 and MWCNTs yields a good agreement with the experiments (Table 4 and Fig. 6). In addition, hybrid GNP-L150/MWCNTs nanocomposites exhibited α=1.22 and η=0.27, whereas for single inclusions α=1.33 and η=0.27 were obtained (Fig 6 (b)). The lower value of α for hybrid GNPs/MWCNTs based nanocomposites proves the better dispersion in PU matrix, which would contribute to thermal conductivity enhancement by forming a conductive 3D structure. On the other hand, the higher ratios of GNPs in hybrid systems weaken the TC of nanocomposites. Excess loading of GNPs leads to the shortage of MWCNTs that cannot effectively bridge the adjacent GNPs and thus results in the breakdown of the 3D conductive network. Consequently, the remarkable synergistic effect that exhibits the best performance can be reached in an optimum ratio of GNPs/CNTs hybrids.

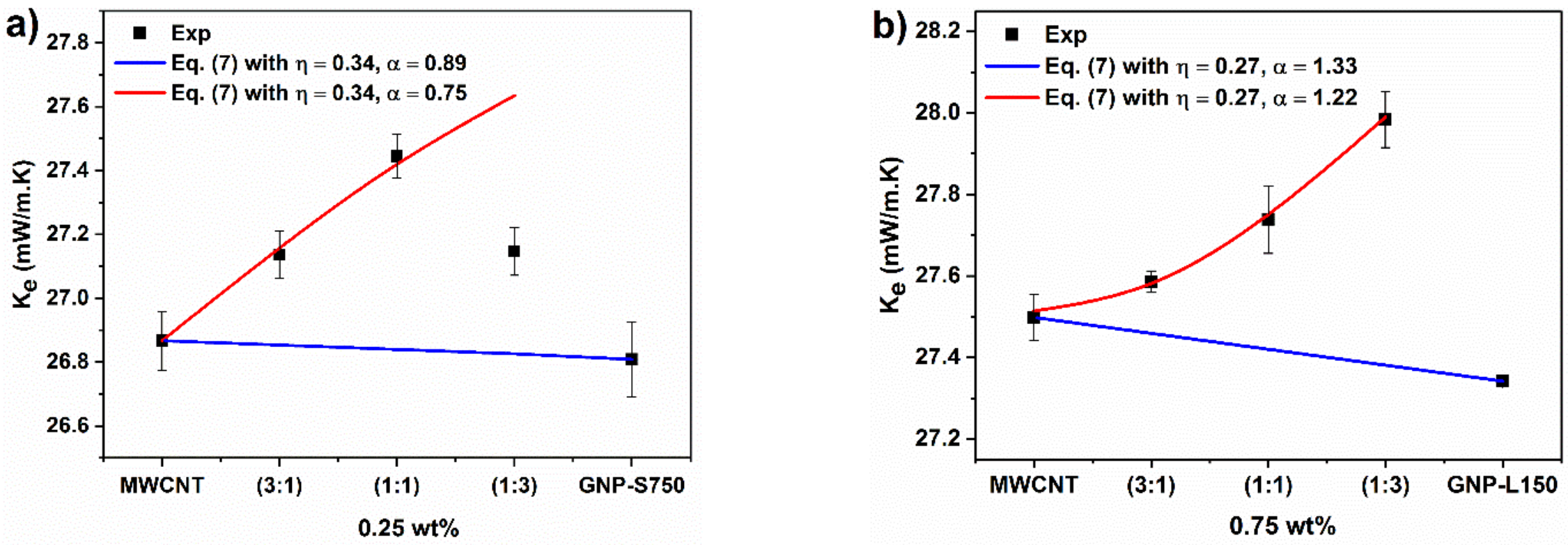

Fig. 6. Model predictions (Eq. 7) for the effective thermal conductivity ($K_e$) of hybrid (a) GNP-S750/MWCNTs (at 0.25 wt%) and (b) GNP-L150/MWCNTs (at 0.75 wt%) based PU nanocomposites compared with the experimental data.

## Conclusion

The thermal conductivity of PU nanocomposites comprising MWCNTs and three types of GNPs was studied by experimental and theoretical approaches. By combining MWCNTs and GNPs a synergistic thermal conductivity improvement was achieved, which surpasses the performance of individual carbon nanofillers in PU matrix. The optimization of hybrid nanofillers is essential to maximize the synergistic effect and the thermal conductivity improvement of the nanocomposites. Synergistic thermal conductivity enhancement using MWCNTs and GNP-S750 with a smaller flake size and higher specific surface area was achieved at low nanofillers contents (0.25 wt%) which is vital for reducing the viscosity and improving the processability of the nanocomposites. In addition, the synergism can be seen at higher nanofiller content (0.75 wt%) for GNP-L150 with a larger flake size. The hybrid nanofillers can effectively enhance the thermal conductivity so that 1D MWCNTs interconnected adjacent 2D GNPs to form a more efficient 3D thermally conductive network. As an interesting outcome, the thermal conductivity of MWCNTs/GNP-S750 (1:1) hybrid nanocomposite at 0.25 wt% is approximately equivalent to the thermal conductivity of the individual GNPs and MWCNTs at 0.75 wt%. Therefore, it can be concluded that the fabrication of hybrid nanocomposites is a cost-effective method with higher processability through declining required nanofillers content. An analytical study on the thermal conductivity of single and hybrid nanocomposites by incorporating the straightness factor and percolation effect as fitting parameters is presented, in which an outstanding agreement between the modified model predictions and experimental results were achieved. The results showed that the straightness of nanofillers profoundly affected the thermal conductivity of nanocomposites and larger GNPs tend to be wavier or less straight in the polymer matrix so that the calculated straightness ratio reduced by enhancing the flake size of GNPs.

## Acknowledgments

The authors are grateful to Prof. Dr. Yakup Erhan Boke for thermal conductivity measurements.

We would like to thank Mr. Osman Celebi for assistance with the mold and samples preparation.